\documentclass{article}

\usepackage[english]{babel} 				
\usepackage{graphicx}					
\usepackage{vmargin}						
\usepackage{authblk}   
\usepackage{hyperref}						
\usepackage{subfig}							
\usepackage{amsmath}				    	
\usepackage{amsfonts}
\usepackage{amssymb}
\usepackage{amsthm}	
\usepackage{caption} 
\usepackage{dsfont}  						
\usepackage{fancyhdr,fancybox} 			   
\providecommand{\keywords}[1]{\textbf{\textit{Key Words:  }} #1}
\providecommand{\msccodes}[1]{\textbf{\textit{MSC2010:  }} #1}
\usepackage{setspace}
\theoremstyle{definition}
\newtheorem{Theorem}{Theorem}[section] 
\newtheorem{Definition}[Theorem]{Definition}
\newtheorem{Assumption}[Theorem]{Assumption}
\newtheorem{Lemma}[Theorem]{Lemma}
\newtheorem{Proposition}[Theorem]{Proposition}
\newtheorem{Remark}[Theorem]{Remark}

\newcommand{\R}{\mathbb{R}}

\newcommand{\E}{\mathbb{E}}

\newcommand{\p}{\mathbb{P}}

\author{Christof Henkel\thanks{henkel.christof@gmail.com}}

\affil{Mathematisches Institut der Ludwig-Maximilians-Universit\"at M\"unchen\\
  Theresienstrasse 39, 80333 M\"unchen, Germany\\}
\date{September 2016}
\title{From quantum mechanics to finance: Microfoundations for jumps, spikes and high volatility phases in diffusion price processes}

\begin{document}

\maketitle
\begin{abstract}
We present an agent behavior based microscopic model that induces jumps, spikes and high volatility phases in the price process of a traded asset. 
We transfer dynamics of thermally activated jumps of an unexcited/ excited two state system discussed in the context of quantum mechanics to agent socio-economic behavior and provide microfoundations. After we link the endogenous agent behavior to price dynamics we establish the circumstances under which the dynamics converge to an It{\^o}-diffusion price processes in the large market limit. 
\end{abstract}

\vfill 

\\
\msccodes{60F05, 60J60, 60K35, 91D30, 91B69, 91B80}
\renewcommand{\thepage}{\arabic{page}}

\setlength{\skip\footins}{10mm} 
\thispagestyle{empty} 
\newpage
\setcounter{page}{1} 

\section{Introduction and Methods}
\subsection{Introduction}
In 1900 the french mathematician Louis Bachelier suggested to use brownian motion to model price fluctuations at the paris stock market (see Bachelier \cite{bach1900}) and so laid the foundations of modern financial mathematics. Through the last century, especially since the introduction of the Black-Scholes model in 1970 (see Black and Scholes \cite{bla1973}), It\^o diffusion processes became the standard tool for modeling and pricing of financial assets. On the other hand people are aware that the single asset price process is a macroscopic result of many microscopic factors like for example individual traders behavior. So it is not surprising that many and various probabilistic models have been invented to study not only the individual agents trading behavior but also their interaction. 

In F{\"o}llmer and Schweizer \cite{foe1993} as well as in Horst \cite{hor2005} stock prices are modeled in discrete time as sequence of temporary equilibria. Those emerge as a consequence of simultaneous matching of supply and demand of several agents. Bayraktar, Horst and Sircar \cite{bay2006},\cite{bay2007}, to account for this asynchronous order arrival, as well as Horst and Rothe \cite{hor2008} use the mathematical framework of queuing theory earlier examined by Mandelbaum, Pats et al. \cite{man1998a} and Mandelbaum, Massey and Reiman \cite{man1998b} for their models. Also the model explained in Lux \cite{lux1995} and Lux \cite{lux1997} takes asynchronous order arrivals into consideration by using a so called market maker who matches supply and demand and alters the price accordingly. Thereby the individual agents behavior is dependend on his opinion. Further Opinion-based models range from binary (e.g. F{\"o}llmer \cite{foe1974}, Arthur \cite{art1994}, Orl{\'e}an \cite{orl1995}, Weisbuch and Boudjema \cite{wei1999} and Sznajd-Weron and Sznajd \cite{szn2000}) to opinions from a continuous spectrum, which are used, for example, to describe large social networks or ratings (see Duffant et al. \cite{def2000}, G{\'o}mez-Serrano, Graham and LeBoudec \cite{gom2012} or Weisbuch, Deffuant and Amblard \cite{wei2005}). However the characteristics and the interaction of the agents are described in the respective model, it mostly can be classified, in a wider sense, as interacting objects with assigned states forming a link to other scientific fields. Especially the application of methods and dynamics derived from physics, captured under the term of econophysics, is becoming increasingly popular and also this paper builds a bridge from physics to socio economics and finance. 

We use an extented version of the microscopic market model for diffusion price processes of Pakannen \cite{pak2010}, which is presented in Henkel \cite{hen2016} and apply dynamics of a quantum system from Bauer, Bernard and Tilloy \cite{bau2015}, respectively Tilloy, Bauer and Bernard \cite{til2015}, to agents social behavior.\\ We transfer statistical properties observed in thermally activated jumps in a quantum system to the average dynamics of market participants and induce the same in diffusion price processes in the large market limit. As such we provide a microscopic explanation for jumps in asset price processes as oberserved in A{\"i}t et al. \cite{ait2015} without using a jump process, as for example employed by Deng \cite{deng2000}. Additionally our dynamics induce spikes and high volatility phases that have also present in various price processes (e.g. Ham \cite{ham1996}).

We structure the content as following. In the first section we set the general market framework as a pool of interacting agents. To describe their interactive behavior we assign each agent heterogeneously an excitement state similar to the excitement of a two level quantum system as in Bauer, Bernard and Tilloy \cite{bau2015} and express the overall market excitement as the distribution of those excitement states. Furthermore we state conditions under which, as a mean-field like result, the overall market excitement can be expressed as a single diffusion process in the large market limit. Then, with defining the agents propensity to trade and by specifying the impact on the asset price we link the endogenous market dynamics to the asset price movement. We specify conditions under which, in a large market, the asset price development can be approximated by a diffusion price process and conclude with a proposition summarizing the diffusion approximation.

\subsection{Methods}
The model presented in this paper was implemented in the statistical programming language R (Ihaka and Gentleman \cite{iha1996}) in order to simulate the related distributions and stochastic processes and to illustrate trajectories. Especially functions, which are presented as solutions of stochastic differential equations without provision of a closed analytic form, are illustrated using an implementation of the Euler-Maruyama-scheme (see e.g. Glasserman \cite{gla2003}). All R-scripts are available on request.

\section{Endogenous dynamics}
Before we link the agents behavior to an asset price, we specify all model components, that directly affect the interaction between market participants. Let $\mathbb{A}_n = \{1, ..., n \}, \ n \in \mathbb{N}$ be a finite set of agents.\\ Following Bauer, Bernard and Tilloy \cite{bau2015} we consider a state space $S=\{s_1,s_2 \} = \{0,1\}$, where $s_1 =0$ represents an "unexcited" state and $s_2=1$ an "excited" state. The vector of all individual states takes values in the \textit{configuration space} \mbox{$C := S^{\mathbb{A}_n} = \{x = (x^a)_{a=1}^n, \ x^a \in S\}$}.
During the time $t \in [0, \infty)$ each agent can consider a state transition. The time of the k-th consideration is nominated by $T_k \geq 0$, $k \in \mathbb{N}$. The action times are described later in detail, however we use the terminology to describe the development of the states within discrete time in the following definition.
The state of agent a at time $T_k$ is defined as $x^a_{T_k} \in S$. We capture the development of agent a's state by the process $(x_k^a)_{k \in \mathbb{N}} := (x^a_{T_k})_{k \in \mathbb{N}}$ and the development of all agents states by the n-dimensional process $(x_k)_{k \in \mathbb{N}} = (x_k^a)_{k \in \mathbb{N},a \in \mathbb{A}_n}$. We assume that the vector of initial states is distributed following some n-dimensional distribution function, in particular $x_0 \sim F_{x_0}^n$, and assign to each agent $a \in \mathbb{A}_n$ an initial state $x_0^a \in S$.

\begin{Definition}[Market excitement]\ \\
We measure the proportion of excited agents in the market at time $T_k$ by the \textit{market excitement}

\begin{equation}
\overline{M}_k = \frac{1}{n} \sum_{a =1}^n \mathds{1}_{\{1\}}(x_k^a), k \in \mathbb{N}.
\end{equation}

Additionally, we denote the initial distribution of the market excitement resulting from $F_{x_0}^n$ as $F_{\overline{M}_0}^n$. Note that, by construction, $\overline{M}_k$ is the average excitement of all agents and a probability measure on $C$.
\end{Definition}

\begin{Definition}[Endogenous market history]\label{Endogenuous_market_history}
 \ \\
We capture all endogenous information up to $T_k$ in the \textit{endogenous market history}, which is given by the sigma algebra $\mathcal{G}_k := \sigma (T_i, A_i,\overline{M}_i, i \leq k)$. Here the tupel $(T_k, A_k, \overline{M}_k), k \in \mathbb{N}$ represents agent $A_k$ who acts at $T_k$ and the resulting market excitement $\overline{M}_k$. We assume that only one agent changes his state at a specific point in time. Although this assumption seems rather strong, it is reasonable as transitions are performed in continuous time and are unlikely to happen at the same time.
\end{Definition}

\begin{Definition}[Transition intensity]\label{Def_transition_intensity}\ \\
In order to heterogeneously specify the agents tendency to consider a state transition we assign to each agent a \textit{transition intensity} $\mu_a = n \gamma^2_a$, that is an agent dependent constant $\gamma_a^2 \in \R^+$ times the number of agents participating in the market. Next, we define the probability that it is agent a who wants to reevaluate his state. Heuristically we weight the individual transition intensity by the sum of all trading intensities, which we call \textit{aggregated transition intensity} and denote by $\mu_{\mathbb{A}_n} = \sum^n_{a=1} \mu_a $. In summary,
\begin{equation}\label{prob_agant_a}
\p (A_k = a |\mathcal{G}_{k-1}) = \frac{\mu_a}{\mu_{\mathbb{A}_n}} = \frac{\gamma_a^2}{\sum_{\hat{a}=1}^n \gamma_{\hat{a}}^2}.
\end{equation}
\end{Definition}

Next, we characterize the state transition laws, i.e. the probability that agent a changes from excited to unexcited and vice versa, given that he is the one that considers a state change, before we consequentially derive the dynamics of the market excitement. 

\begin{Definition}[Transition probabilities]
We use the following notation for the two individual state transition probabilities.
\begin{equation}\label{trans_prob1}
\Pi^{1,2}_{n,a}(\overline{M}_{k-1}) = \beta_a \frac{p_a}{2 \gamma^2_a n} + \frac{\eta_a  h^{1,2}(\overline{M}_{k-1}) }{2}
\end{equation} 
\begin{equation}\label{trans_prob2}
\Pi^{2,1}_{n,a}(\overline{M}_{k-1}) = \beta_a \frac{1-p_a}{2 \gamma^2_a n} + \frac{\eta_a h^{2,1}(\overline{M}_{k-1})}{2},
\end{equation} 
where $\beta_a,p_a,\eta_a \in [0,1]$ are agent dependent constants and

\begin{equation}\label{Def_h}
h^{i,j}(y) = (1-y)^i y^j, \ \ y \in [0,1], i \neq j \in \{1,2\}.
\end{equation}

We capture all state transition probabilities per agent in a transition matrix, i.e. we define
\begin{equation}
\Pi_{n,a}(\overline{M}_{k-1}) = \begin{pmatrix}  1-\Pi_{n,a}^{1,2} & \Pi_{n,a}^{1,2}   \\  \Pi_{n,a}^{2,1} & 1-\Pi_{n,a}^{2,1}  \\ \end{pmatrix} (\overline{M}_{k-1}).
\end{equation}
\end{Definition}

\begin{Remark}
We choose this explicit form of transition probabilities presented in Equations (\ref{trans_prob1}) and (\ref{trans_prob2}) for the following reasons. The first part of the sum models individual intrinsic disposition for excitement. Thereby $p_a$, respectively $1-p_a$, captures the distance from agents $a$'s actual excitement state and his individual preference.\footnote{Note that $\Pi^{1,2}_{n,a}$ is only relevant for unexcited agents ($x^a_{k-1} = 0$) and analogously $\Pi^{2,1}_{n,a}$ only for agents with $x^a_{k-1} = 1$. Hence the simplified form in Equations (\ref{trans_prob1}) and (\ref{trans_prob2}).} So we heuristically reflect a higher drive to transition when the distance to the personal preference is large. Apart from autonomous behavior we also want to model influence of other agents on the individuals excitement state (In order to study later on a possible impact on the resulting price process). For this we use the second addend which takes into account the average excitement of all agents and thus models herd behavior. By the choice of the form of  $h^{i,j}$ (Equation (\ref{Def_h})), an unexcited agent ($x^a_{k-1} = 0$) has a higher probability to transition if the market excitement is large. Analogously, the transition probability to become unexcited is bigger when the market excitement is low, that is, if the majority of agents is unexcited. Besides being simple and symetric, $h^{i,j}$ also induces the same dynamics in the large market limit as a continuous measurement of the quantum system\footnote{See Bauer, Bernard and Tilloy \cite{bau2015}.}, which further supports the choice. We weight the two aspects, that is autonomous behavior and heteronomy, individually per agent by constants $\beta_a$ and $\eta_a$. Moreover we scale the first part by $\mu_a$ to get a well designed probability measure and to model the increasing importance of herding when the market is large.
\end{Remark}

Since the average opinion $\overline{M}_k$ can from time $T_k$ to time $T_{k+1}$ either change by $\pm \frac{1}{n}$ or stay unchanged, it has values on the $n+1$ valued lattice $\mathbb{L}$ from 0 to 1, viz.
\begin{equation}
\overline{M}_k \in \mathbb{L}, \ \forall k \geq 0, \ \text{with} \ \mathbb{L}:= \left\{0, \frac{1}{n}, \dots, \frac{n-1}{n}, 1  \right\}.
\end{equation}

In summary, $(\overline{M}_k)_{k \geq 0}$ is a Markov chain on $\mathbb{L}$ with value dependent transition probabilities, which are stated in the next Lemma.

\begin{Lemma}[Discrete market excitement dynamics]\label{discrete_market_dynamics}\ \\
The probability that any excited agent becomes unexcited and therefore that $\overline{M_k}$ decreases by $1/n$ is given by 
\begin{equation}
\begin{split}
\p \left(\overline{M}_k - \overline{M}_{k-1} = -\frac{1}{n} \Big| \mathcal{G}_{k-1}\right) 
&= \frac{\sum^n_{a = 1} \left( \beta_a (1-p_a)\overline{M}_{k-1} + n \eta_a \gamma_a^2 (1-\overline{M}_{k-1})^2 \overline{M}_{k-1}^2\right)}{2n \sum_{a=1}^n \gamma_a^2} 
\end{split}
\end{equation}

and similarly the probability that the market excitement increases by $1/n$ is given by
\begin{equation}
\begin{split}
\p \left(\overline{M}_k - \overline{M}_{k-1} = \frac{1}{n} \Big| \mathcal{G}_{k-1}\right) 
&= \frac{\sum^n_{a = 1} \left(\beta_a p_a(1-\overline{M}_{k-1}) + n \eta_a \gamma_a^2 (1-\overline{M}_{k-1})^2 \overline{M}_{k-1}^2\right)}{2n \sum_{a=1}^n \gamma_a^2} 
\end{split}
\end{equation}
\begin{proof}
Using the fact that $\overline{M}_{k-1}$ is a probability measure on $C$ as well as the representation defined in Equation (\ref{prob_agant_a}),(\ref{trans_prob1}) and (\ref{trans_prob2}) the lemma follows from basic calculations.
\end{proof}
\end{Lemma}

In order to embed the Markov chain $(\overline{M}_k)_{k \geq 0}$ homogeneously in continuous time and thus describing the market excitement by a time homogeneous Markov process, we further characterize the points in times at which the agents decide to make a transition. 

\begin{Definition}[Transition times]\label{Def_intra_transition}\ \\
The \textit{transition times} $(\tau_k)_{k \geq 1}$ are defined as $\tau_k := T_k - T_{k-1}, k \geq 1.$\\
Since we want the transition times to be memory-less for the sake of simplicity, i.e.
\begin{equation}
\p (\tau_k > t +h | \tau_k > h, \mathcal{G}_{k-1}) = \p(\tau_k > t | \mathcal{G}_{k-1}), \ t,h \geq 0.
\end{equation}
the transition times are assumed to be exponentially distributed. Heuristically we assume that the rate of the exponential distribution is given by the aggregated transition intensity, i.e.
\begin{equation}
\p (\tau_k \in [0,t]|\mathcal{G}_{k-1}) = 1 - e^{-n t\sum_{a=1}^n \gamma_a^2 }, \ t \geq 0,
\end{equation}
\end{Definition}
\newpage
\begin{Definition}[Market excitement index]\ \\
After fixing $T_0 = 0$ we can define the  \textit{market excitement index} via
\begin{equation}
Q^n_t := \sum^\infty_{k=0} \overline{M}_k \mathds{1}_{[T_k,T_{k+1})}(t), \ t \geq 0.
\end{equation} 
\end{Definition}

Note that, by construction, $Q^n_t$ is c\'{a}dl\'{a}g and a well defined time homogeneous pure jump type Markov process. Its existence is stated in the following lemma. The basis of the lemma builds the synthesis theorem (e.g. Theorem 12.18 of Kallenberg \cite{kal2002}), which embeds a discrete Markov chain into continuous time using exponentially distributed waiting times.  

\begin{Lemma}[Existence]\label{Existence_Mk}\ \\
	There exists a probability space $(\Omega, \mathcal{F}, \mathbb{P})$ in which $(Q^n_t)_{t \in [0, \infty)}$ is a time homogeneous pure jump Markov process with rate kernel 
	
	\begin{equation}
	K_n(q,s) := n\sum_{a=1}^n \gamma^2_a k_n(q,s), 
	\end{equation}
	
	The transition kernel $k_n(q,s),\ s \in \left\{-\frac{1}{n}, 0 ,\frac{1}{n}  \right\}$ is a regular version of the conditional distribution $\mathbb{P}(\overline{M}_1 - \overline{M}_0 = s | \overline{M}_0 = q)$, which is given by Lemma \ref{discrete_market_dynamics}.
	\begin{proof}
	By the construction of the Markov chain $(\overline{M}_k)_{k \geq 0}$ and the assumption made in Definition \ref{Def_intra_transition} the synthesis theorem (e.g. Theorem 12.18 of Kallenberg \cite{kal2002}) states that $(Q^n_t)_{t \in [0, \infty)}$ is a pure jump-type Markov process and also gives the rate kernel. Time homogeneity is given by the recursive definition of $(\overline{M}_k)_{k \geq 0}$ (see e.g. Proposition 8.6 of Kallenberg \cite{kal2002}) as the transition matrix $\Pi_{n,a}$ is independent of time.
	\end{proof}
\end{Lemma}

Although the heterogenous agents are allowed to have individual parameters, in order to ensure a convergence to a mean-field like single equation in the large market limit, the scaled parameters should tend to their mean when the number of market participants goes to infinity. This we summarize in the next Assumption.

\begin{Assumption}\label{Assumption_Q}\ \\
We assume
\begin{enumerate}
\item $F_{\overline{M}_0}^n \xrightarrow{n \to \infty} F_{\overline{M}_0}$
\item $\sum^n_{a=1} \frac{\gamma_a^2}{n} \xrightarrow{n \to \infty}  \gamma^2, \ \sum^n_{a=1} \frac{\eta_a}{n} \xrightarrow{n \to \infty} \eta$
\item $\sum^n_{a=1} \frac{\beta_a}{2n} \xrightarrow{n \to \infty} \beta$ and $\sum^n_{a=1} \frac{p_a}{n} \xrightarrow{n \to \infty} p$
\end{enumerate}

for some constants $\beta, \gamma, \eta, p$ and $F_{\overline{M}_0}$ being a probability distribution.
\end{Assumption}

Now, we are ready to state the large market limit for the market excitement index. 

\begin{Proposition}[Large market approximation]\label{large_market_approx_Q}\ \\
	If Assumption \ref{Assumption_Q} holds, then
	\begin{equation}
	(Q^n_t)_{t \in [0, \infty)} \xrightarrow{\mathcal{L}} (Q_t)_{t \in [0, \infty)} \ in \ D_{[0,1]}[0, \infty),
	\end{equation}
	with $(Q_t)_{t \in [0, \infty)}$ being the unique strong solution of the stochastic differential equation (SDE)
	\begin{equation}\label{SDE_Q_t}
	dQ_t = \beta (p-Q_t)dt + \gamma \sqrt{\eta} (1-Q_t) Q_t dB_t, \ \ Q_0 = \theta ,
	\end{equation}
	where $(B_t)_{t \in [0, \infty)}$ is a one dimensional standard Brownian motion, which is independent of $\theta \sim F_{\overline{M}_0}$.
	\begin{proof}
	See Appendix \ref{proof_prop_large_market_Q}.
	\end{proof}
\end{Proposition}

To illustrate properties of the large market limit $Q_t$ we show two trajectories of the solution of Equation (\ref{SDE_Q_t}) for $p=0.6$, $\eta = \beta= 1$ below in Figure \ref{Q_t_gamma1} and Figure \ref{Q_t_gamma10}. The appearance of the process strongly depends on the value of $\gamma$. For a small $\gamma$, as shown in Figure \ref{Q_t_gamma1} with $\gamma = 1$, the market excitement index moves towards an equilibrium at the constant $p \in [0,1]$, which is given as the mean of the individual preference level $p_a$ (see Assumption \ref{Assumption_Q} 3.). Setting a high value of $\gamma$ (see Figure \ref{Q_t_gamma10}, where $\gamma = 10$) the market excitement index is pulled towards the two states $s_1= 0$ and $s_2 = 1$ with jumps and spikes in between. 
Although the separation into two cases (i.e. one or two equilibria) by the value of $\gamma$ is not obvious from the underlying SDE (\ref{SDE_Q_t}), it is expected from the microscopic modeling. In the individual transition probabilities (see Equation (\ref{trans_prob1}) and (\ref{trans_prob2})) $\gamma_a$ scales down the agents individual autonomy and hence represents the exposure to herding. Respectively, $\gamma$ reflects the average herding intensity. It was already shown in Lux \cite{lux1995}, with a similar setup, that minor herding behavior results in a single equilibrium, while strong herd behavior results in two temporary equilibria with phase transitions. In our model the two states $s_1$ and $s_2$ serve as the two temporary equilibria, where jumps represent phase transitions and the spikes imply unsucessful jump attempts. Note that the probability to be in the equilibrium $s_1$ is equal to $p$ (see Tilloy, Bauer and Bernard \cite{til2015}) and that the behavior is similar to Kramer's double well potential (See Kramer \cite{kra1940}).

\begin{figure}[h!]
{\centering
\includegraphics[width=1\textwidth, height=4.5cm]{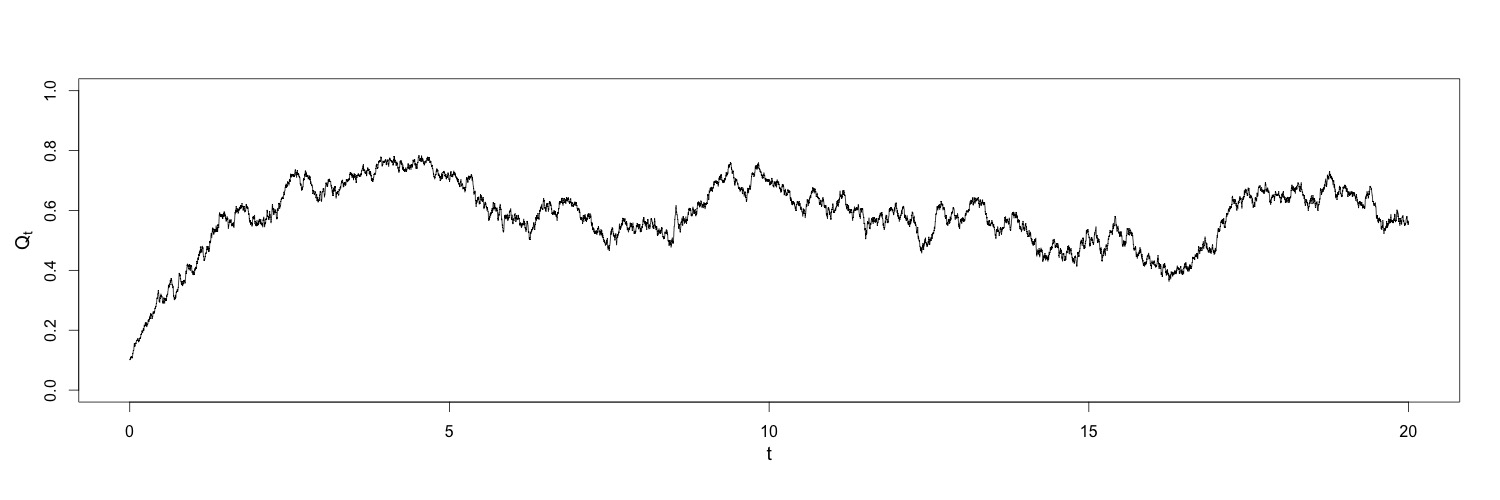}
\captionof{figure}{$Q_t$ for $\gamma = 1$}\label{Q_t_gamma1}
}
\end{figure} 

\begin{figure}[h!]
{\centering
\includegraphics[width=1\textwidth, height=4.5cm]{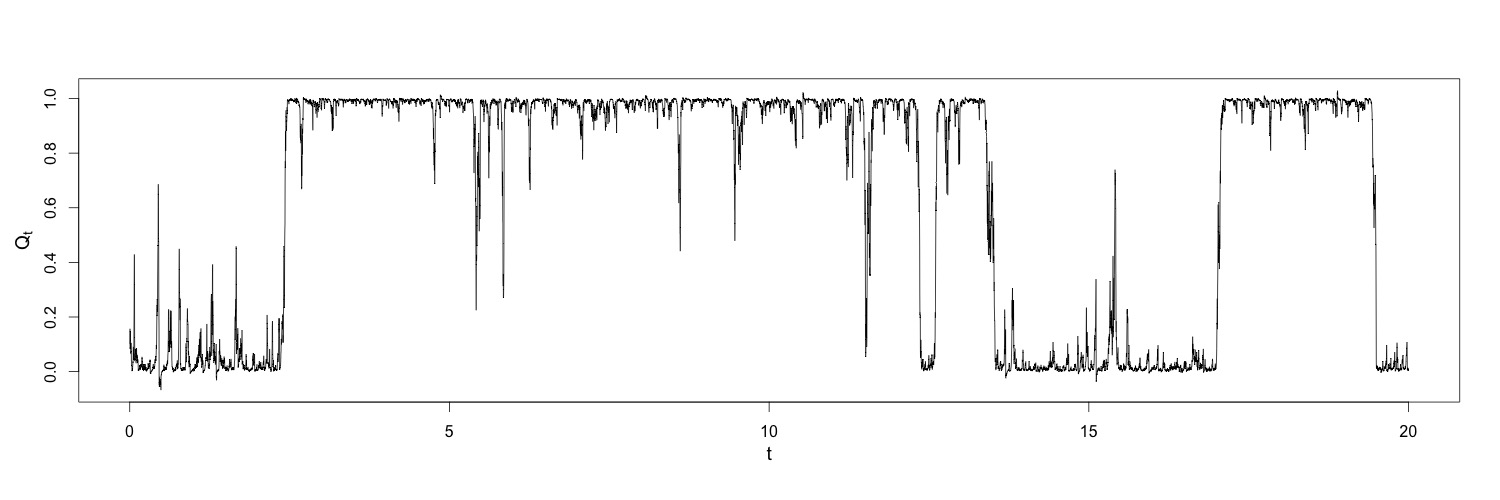}
\captionof{figure}{$Q_t$ for $\gamma = 10$}\label{Q_t_gamma10}
}
\end{figure} 

\begin{Remark}\ \\
Although the SDE presented in Equation (\ref{SDE_Q_t}) is exactly the same as in Bauer, Bernard and Tilloy \cite{bau2015}, it arises differently. In Bauer, Bernard and Tilloy \cite{bau2015} the SDE is the result of a transition from discrete to continuous time in the measurement of a single quantum system, while in our model the SDE is induced by the number of interacting objects tending to infinity.
Also the origin of the spikes and jumps in the large market limit is quite different. While in Bauer, Bernard and Tilloy \cite{bau2015} the jumps and spikes are a result of tight monitoring, in our model both is a result of agents social behavior. Here the root cause of the jumps as well as of the spikes is the agent's individual exposure to mutual interference $\gamma_a$, that is his herding behavior. When the average herding behavior is strong (i.e. $\gamma$ is large), the fast contagion of agents excitement leads to hypes. We show in Section \ref{sec_pd} that these hypes result in phases of high price volatility, where transition to a hype is indicated by a jump in the price process and spikes are unsuccessful  jump attempts.
\end{Remark}

\section{Price dynamics}\label{sec_pd}
In this section we link the endogenous dynamics of the previous section with dynamics of an asset price. We assign to each trader an individual trading intensity quantifying his propensity to trade and a excess demand function which characterizes the quantity of shares bought or sold. We then define a pricing rule according to which the number of bought or sold shares impacts the price.
To further show the flexibility of our model, we introduce an additional group of traders called fundamentalists. Last are characterized by basing their behavior on the difference between actual price and a \textit{fundamental value} $F \in \mathbb{R}$. In particular, when the price is below (above) $F$, they consider the asset cheap (expensive) and want to buy (sell). We assume the fundamentalists are homogeneous, viz. $F$ is common, and the fundamental value is time-invariant. 

Let $\mathbb{A}_n = \{1, ..., n \}, \ n \in \mathbb{N}$ be the set of agents of which a fixed subset $\mathbb{F}_n \subseteq \mathbb{A}_n$ with $|\mathbb{F}_n| = k_n \in \{0, ..., n\}$ are fundamentalists and the rest are noise traders. We denote the portion of fundamentalists with $\phi_n = k_n / n$. We assume that fundamentalists are unexcited and have no desire to change their state, i.e. $\forall a \in \mathbb{F}_n: \ x_0^a = 0, \ \beta_a = \eta_a = 0$. Note that, alternatively we could have introduced fundamentalists as an additional state. However, the current setup illustrates the flexibility arising from heterogeneous transition probabilities.

We assume that any change of the market is a direct consequence of agents behavior. The behavior is given by actions that can either be the change of state or a trade of the asset. We index each of these actions by $k \in \mathbb{N}$ 
and extent the endogenous market history defined in Definition \ref{Endogenuous_market_history} with the trading behavior.

\begin{Definition}[k-th action, market history]
 \ \\
The \textit{k-th action} is characterized by the tupel $(\tilde{T}_k, A_k, P_k, \overline{M}_k, B_k), k \in \mathbb{N}$, where $\tilde{T}_k$ is the time when the action occurs,
$A_k \in \mathbb{A}_n$ is the acting agent at time $\tilde{T}_k$ and $B_k \in \{0,1\}$ is an \textit{action indicator} whether the agent trades ($B_k = 1$) or changes his state ($B_k = 0$).  $P_k$ is the \textit{price} per share and $\overline{M}_k$ the above mentioned market excitement. All information is captured in the \textit{market history}, which is given by $\tilde{\mathcal{G}}_k := \sigma (\tilde{T}_i, A_i, P_i,\overline{M}_i,B_i , i \leq k)$. 
\end{Definition}

Similar to the transition intensity (Defintion \ref{Def_transition_intensity}), we set a function, that captures each agents propensity to trade the asset and call the sum of both \textit{action rate}. 
\newpage
\begin{Definition}[Trading intensity, action rate]\ \\
We assume that the agents propensity to trade is given by the \textit{trading intensity}
\begin{equation}
\lambda_a= \bar{\lambda}_a + C_e x^a_{k-1},
\end{equation}
with $\bar{\lambda}_a \in \R^+$ being an agent dependent basic trading intensity and $C_e \in \R^+$ a positive constant, which reflects the positive impact of excitement on the propensity to trade.\\
Moreover we introduce the \textit{action rate} for each agent with
\begin{equation}
\nu_a = \mu_a + \lambda_a.
\end{equation}
The aggregated action rate is then given as
\begin{equation}
\nu_{\mathbb{A}_n} = \sum_{a=1}^n \nu_a = C_e \overline{M}_k + \sum_{a=1}^n (n\gamma_a^2 + \bar{\lambda}_a)
\end{equation}
\end{Definition}

Analogue to Equation (\ref{prob_agant_a}) we identify the acting agent and the related action by weighting the respective intensity functions. 

\begin{Definition}(Acting probabilities)\ \\
The probability, that agent a trades at $\tilde{T}_k$ is defined as 
\begin{equation}\label{prob_trade}
\p (A_k = a, B_k = 1|\tilde{\mathcal{G}}_{k-1}) = \frac{\lambda_a }{\nu_{\mathbb{A}_n}}.
\end{equation}

Similarly, we define the probability, that agent a changes his state by

\begin{equation}\label{prob_trade2}
\p (A_k = a, B_k = 0|\tilde{\mathcal{G}}_{k-1}) =\frac{\mu_a }{\nu_{\mathbb{A}_n}}.
\end{equation}

Moreover the probability that the k-th action is a state transition is set as

\begin{equation}
\p (B_k = 0|\tilde{\mathcal{G}}_{k-1}) = \sum_{a=1}^n \frac{\mu_a }{\nu_{\mathbb{A}_n} } = \frac{\mu_{\mathbb{A}_n} }{\nu_{\mathbb{A}_n} },
\end{equation}

and analogously the probability that the k-th action is a trade is given by

\begin{equation}
\p (B_k = 1|\tilde{\mathcal{G}}_{k-1}) = \frac{\lambda_{\mathbb{A}_n} }{\nu_{\mathbb{A}_n}} = 1- \p (B_k = 0|\tilde{\mathcal{G}}_{k-1}).
\end{equation}
\end{Definition}

Next we define the traded quantity per agent once he decided to trade. Thereby we differ between noise traders and fundamentalists. While fundamentalists base their excess demand on the difference between the last known price and the fundamental value, noise traders trade according to random signals $(\xi_k)_{k \geq 1}$ which are assumed to be i.i.d. with $\E[\xi_1] = 0$ and $ \sigma_{\xi}^2 := \E[\xi_1^2] < \infty$. Thereby the variance of the traded quantity is determined by the variance of the market excitement $\overline{M}_k$. 

\begin{Definition}[Excess demand function]\ \\
In summary we set the following \textit{excess demand function}
\begin{equation}
e_a(P_{k-1},\overline{M}_{k-1},\xi_k) =
\begin{cases}
\frac{1}{\sqrt{n}} (F - P_{k-1}), & a \in \mathbb{F}_n\\
\xi_k \gamma_a^2 \eta_a \overline{M}_{k-1}^2 (1-\overline{M}_{k-1})^2, &  a \notin \mathbb{F}_n,
\end{cases}.
\end{equation}
\end{Definition}

After an agent decides to trade, the new price at time $\tilde{T}_k$ will be set by a market maker, who is assumed to handle all trades, and is defined by a pricing rule depending on the excess demand of the acting agent and the old price.
We assume that the difference of old and new price is linear in the quantity traded root-scaled by the number of market participants. That is  
\begin{equation}
P_k = r_n(e_a(P_{k-1},\overline{M}_{k-1},\xi_k),P_{k-1}),
\end{equation}

where 

\begin{equation}
r_n(q,x) = x+ \frac{\alpha}{\sqrt{n}} q.
\end{equation}

By construction $(P_k)_{k \geq 0}$ and $(\overline{M}_k)_{k \geq 0}$ are now two interacting Markov chains. In order to embed them homogeneously in continuous time and thus describing the price as well as the character by a time homogeneous Markov process, we further characterize the points in times at which the agents decide to act. 

\begin{Definition}[Intra-action times]\label{Def_intra-action}\ \\
The \textit{intra-action times} $(\tilde{\tau}_k)_{k \geq1}$ are defined as $\tilde{\tau}_k := \tilde{T}_k - \tilde{T}_{k-1}, k \geq 1.$\\
Analogously to Definition \ref{Def_intra_transition}  we assume that the rate of the exponential distribution is given by the aggregated action rate, i.e.
\begin{equation}\label{exponential_waiting_times}
\p (\tilde{\tau}_k \in [0,t]|\tilde{\mathcal{G}}_{k-1}) = 1 - e^{-t\nu_{\mathbb{A}_n}}, \ t \geq 0,
\end{equation}
More precisely, to ensure a sufficient level of independence between the source of randomization and the price as well as market character we need to assume that the Intra-action times $(\tilde{\tau}_k)_{k \geq 1}$ are given by 
\begin{equation}
\tilde{\tau}_k := \frac{\psi_k}{\nu_{\mathbb{A}_n}}, \ k \in \mathbb{N},
\end{equation} 
where $(\psi_k)_{k \geq 1}$ are i.i.d. random variables independent of $(P_k,\overline{M}_k)_{k \geq 0}$ with $\psi_1 \sim Exp(1)$.
\end{Definition}

\begin{Definition}[Price process]\label{def_priceprocess}\ \\
After setting an initial price $P_0 \sim F_{P_0}$ and fixing $\tilde{T}_0 = T_0 = 0$ we can define the \textit{price process} as
\begin{equation}
X^n_t := \sum^\infty_{k=0} P_k \mathds{1}_{[T_k,T_{k+1})}(t), \ t \geq 0.
\end{equation} 

\end{Definition}

We extent Lemma \ref{Existence_Mk} with the Price process in the following.

\begin{Lemma}[Existence]\label{lemma_existence}\ \\
If the preceding Assumptions hold true, then a probability space $(\tilde{\Omega}, \tilde{\mathcal{F}}, \tilde{\mathbb{P}})$ exists,  which carries the model in a way that $(X^n_t,Q^n_t)_{t \in [0, \infty)}$ is a time homogeneous pure jump Markov process with rate kernel 

\begin{equation}\label{rate_kernel}
K_n(x,q,dy,s) := \nu_{\mathbb{A}_n}  k_n(x,q,dy,s),
\end{equation}

where the transition kernel $k_n(x,q,dy,s)$ is a regular version of the conditional distribution\\
$\mathbb{P}(P_1 - P_0 \in dy, \overline{M}_1 - \overline{M}_0 = s | P_0 = x, \overline{M}_0 = q)$, $s \in \{-\frac{1}{n}, 0, \frac{1}{n}\}.$ 
\begin{proof}
Analogous to Lemma \ref{Existence_Mk}.
\end{proof}
\end{Lemma}

Before we can state the large market limit for the market excitement index and the price process, we assume some stability of the proportion of fundamentalists. Additionally we require a mean-convergence of the trading intensities of the fundamentalists as well as of the noise traders.

\begin{Assumption}\label{Assumption_X}\ \\
We assume
\begin{enumerate}
\item $\phi_n \xrightarrow{n \to \infty} \phi$
\item $\frac{1}{n} \sum_{a \in \mathbb{F}_n}^n \bar{\lambda}_a \ \xrightarrow{n \to \infty} \bar{\lambda}_F$
\item $\frac{1}{n} \sum_{a \notin \mathbb{F}_n}^n \bar{\lambda}_a \ \xrightarrow{n \to \infty} \bar{\lambda}_N$
\end{enumerate}

for some constants $\phi \in [0,1], \bar{\lambda}_F \in \mathbb{R}^+$ and $\bar{\lambda}_N \in \mathbb{R}^+$.
\end{Assumption}

Next, we state the SDE whose solution approximates the endogenous dynamics and the price process in a large market, that is a market with many participants.

\begin{figure}[h!]
{\centering
\includegraphics[width=1\textwidth, height=4.4cm]{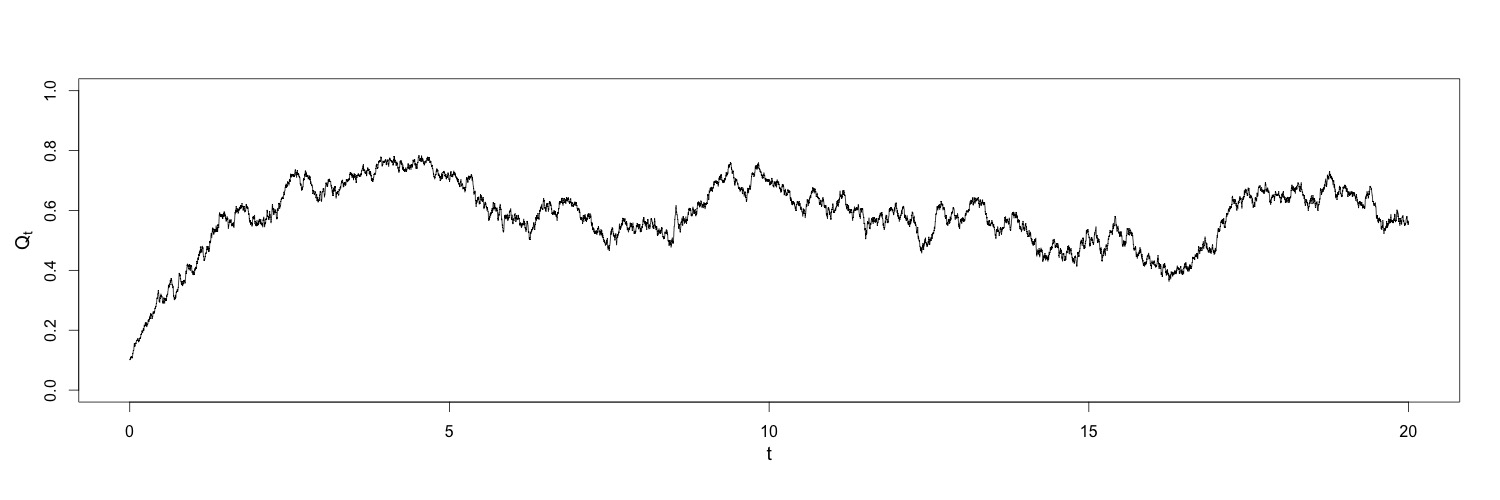}
\captionof{figure}{$Q_t$ for $\gamma = 1$}\label{Q_t_gamma1_2}

}
\end{figure}

\begin{figure}[h!]
{\centering
\includegraphics[width=1\textwidth, height=4.4cm]{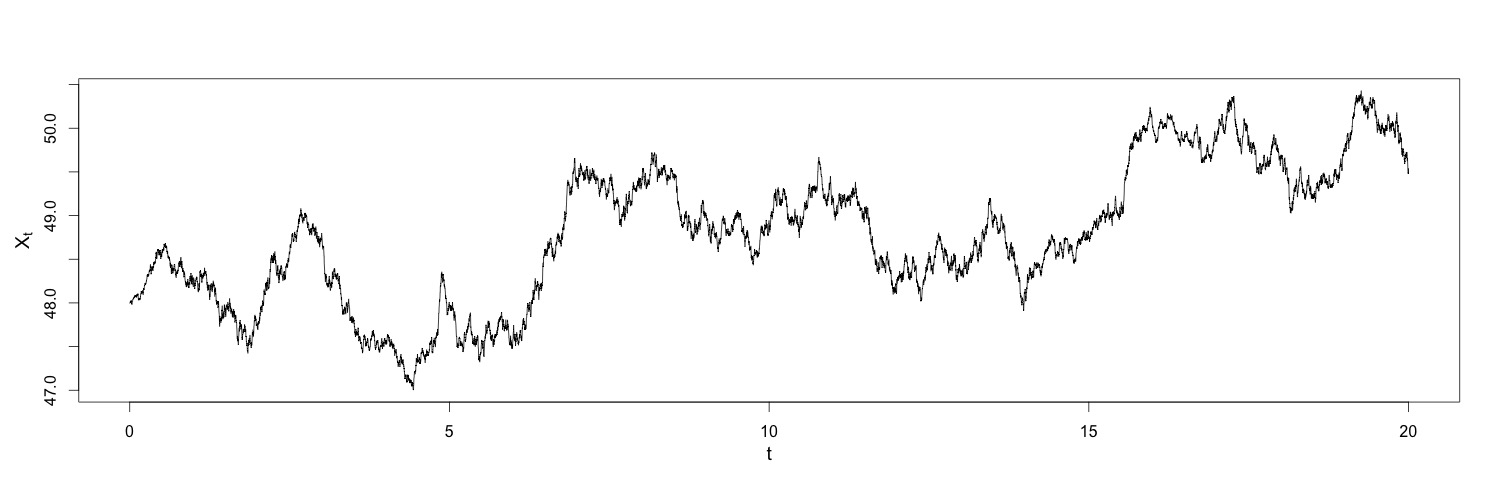}
\captionof{figure}{$X_t$ for $\gamma = 1$}\label{X_t_gamma1}
}
\end{figure}

\begin{figure}[h!]
{\centering
\includegraphics[width=1\textwidth, height=4.4cm]{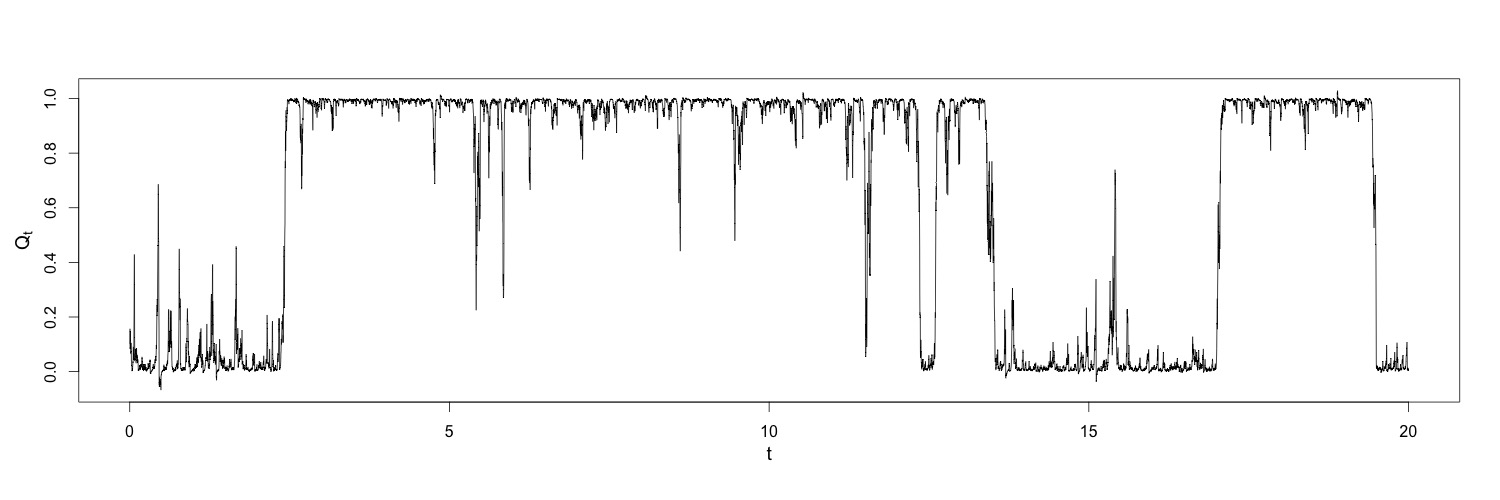}
\captionof{figure}{$Q_t$ for $\gamma = 10$}\label{Q_t_gamma10_2}
}
\end{figure}

\begin{figure}[h!]
{\centering
\includegraphics[width=1\textwidth, height=4.4cm]{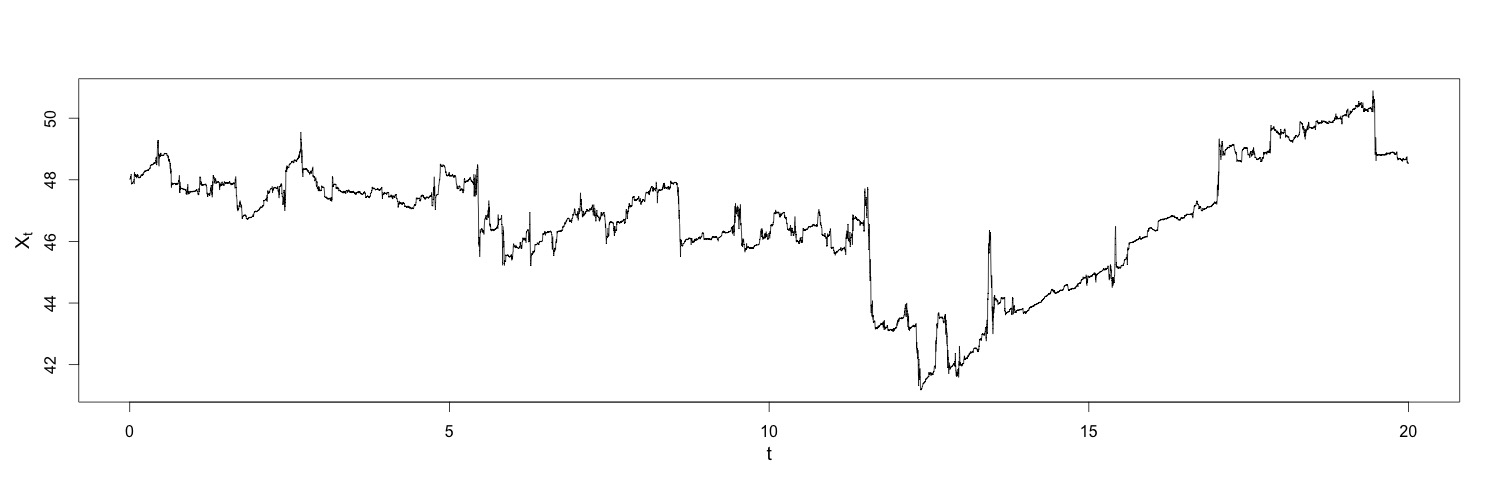}
\captionof{figure}{$X_t$ for $\gamma = 10$}\label{X_t_gamma10}
}
\end{figure}

\begin{Proposition}[Diffusion approximation]\label{Diff_approx}\ \\

If Assumptions \ref{Assumption_Q} and \ref{Assumption_X} hold, then

\begin{equation}
(X^n_t, Q^n_t)_{t \in [0, \infty)} \xrightarrow{\mathcal{L}} (X_t,Q_t)_{t \in [0, \infty)} \ in \ D_{\mathbb{R} \times [0,1]}[0, \infty),
\end{equation}

where $(X_t,Q_t)_{t \in [0, \infty)}$ is the unique strong solution of the SDEs

\begin{equation}\label{SDE1}
\begin{cases}
dQ_t = \beta (p-Q_t)dt + \gamma \sqrt{\eta} (1-Q_t) Q_t dB_t, &Q_0 = \theta\\
dX_t = \bar{\lambda}_F (F-X_t) dt +  \sigma_{\xi} \sqrt{\bar{\lambda}_N + C_e Q_t}\gamma \sqrt{\eta} (1-Q_t) Q_t dW_t, & X_0 = \zeta
\end{cases}
\end{equation}

where $(B_t)_{t \in [0, \infty)}$ and $(W_t)_{t \in [0, \infty)}$ are independent one dimensional standard Brownian motion, $\zeta \sim F_{P_0}$ independent of $W_t$, and $\theta \sim F_{\overline{M}_0}$ independent of $B_t$.
\begin{proof}
See Appendix \ref{proof_diff_approx1}.
\end{proof}
\end{Proposition}
Equation (\ref{SDE1}) summarizes our model in the large market limit. The endogenous behavior is described by $Q_t$ given by the first SDE, which is not depending on the price process $X_t$ and is the same as in the previous section. On the contrary $X_t$ depends on $Q_t$. Not only the volatility coefficient of $Q_t$ reappears in the SDE defining $X_t$, but $Q_t$ also scales the volatility of $X_t$ with the factor $\bar{\lambda}_N + C_e Q_t$. Last leads to high volatility phases when the majority of agents is excited. To illustrate the properties of $(X_t, Q_t)_{t \geq 0}$ we show two trajectories. Thereby we repeat the figures of $Q_t$ from the previous section for readers convenience.\\
In Figure \ref{Q_t_gamma1_2} and \ref{X_t_gamma1} we show the first case with a trajectory of $X_t$ with $F=50$, $\phi = 0.2$, $\delta = 2$ and endogenous dynamics, that is $Q_t$, with parameters $p=0.6$, $\eta = \beta= 1$ and $\gamma = 1$. Driven by 20\% of the agents being fundamentalists, $X_t$ 
 drifts to the fundamental value. Thereby the volatility is rather stable, since $Q_t$ has a single equilibrium at $p=0.6$ and the rest of the volatility coefficient of $X_t$ consists of constants. We illustrate the second case in Figure \ref{Q_t_gamma10_2} and Figure \ref{X_t_gamma10} with the same parameters but setting $\gamma = 10$. There the spikes and phase transitions from $Q_t$ are transferred to the price process and result in spikes and jumps.
 Moreover, s explained above, the phases of temporary equilibrium of $Q_t$ at $s_1 = 1$ comply with high volatility phases of $X_t$, since the factor $ \bar{\lambda}_N + C_e Q_t$ increases the volatility coefficient of $X_t$ when $Q_t$ is large. The intensity of the effect of last is specifically steered by the constant $C_e$.

\newpage

\newpage

\begin{Remark}[Approximation with Poisson Jump Process]
Since the SDE of $Q_t$ in Equation (\ref{SDE1}) is exactly the same as in Tilloy, Bauer and Bernard \cite{til2015}, we can leverage from a result on statistical properties presented there in Proposition 2. The position of spikes and jumps (that is positions of local maxima and minima) can be approximated, when $\gamma$ is large, by two Poisson point processes $N^0_t$ and $N^1_t$ on $[0,1] \times \mathbb{R}^+$ with intensities
\begin{equation}
\begin{cases}
d\Lambda_0 = \beta p dt \left[ \delta (1- N^0_t) dN^0_t + \frac{dN^0_t}{(N^0_t)^2} \right], & N_0^0 = \theta\\
d\Lambda_1 = \beta (1-p) dt \left[ \delta (N^1_t) dN^1_t + \frac{dN^1_t}{(1-N^1_t)^2}\right], & N_0^1 = 1-\theta,
\end{cases}
\end{equation}
where $\delta$ is the delta function.
\end{Remark}

\begin{Remark}
While the statistics of $Q_t$ are discussed in Bauer, Bernard and Tilloy \cite{bau2015}, respectively Tilloy, Bauer and Bernard \cite{til2015}, the statistics of $X_t$ (especially the structure of the spikes) were not studied yet although they are a direct consequence of $Q_t$. One might use the Fokker Planck approximation to approximate the stationary distribution of $X_t^n$, respectively $X_t$, to get more insight. Additionally, in order to study $X_t$, it might also be worth to study the underlying Markov chain $(P_k)_{k \geq 0}$. However both is out of this articles scope.
\end{Remark}

\section{Conclusion and Outlook}
We have proposed a microscopic agent-based model to explain jumps, spikes and high volatility phases in diffusion price processes. Using the mathematical framework of Henkel \cite{hen2016} we set a finite network of heterogenous agents interacting in continuous time. The agents behavior is thereby inspired by the dynamics of excited particles in a quantum system (see Bauer, Bernard and Tilloy \cite{bau2015}). In a second step we link the endogenous dynamics to an asset price process by specifying agents individual trading propensity and excess demand functions together with an overall pricing rule. Furthermore, we showed the conditions under which the average agent excitement as well as the price process converge to a diffusion process when the number of market participants tends to infinity. Since our model induces large market dynamics that are likewise present in the discussion of quantum systems coupled to a thermal bath with continuous monitoring (see Bauer, Bernard and Tilloy \cite{bau2015}) we build a bridge between quantum mechanics and financial mathematics. So we could leverage from the statistical properties of quantum trajectories and apply a result of Tilloy, Bauer and Bernard \cite{til2015} to our asset price model by which the occurring jumps and spikes can be approximated by two Poisson processes.\\
For the sake of simplicity several assumptions have been made, that also show limitations of the model. For instance, the missing feedback of the price process on the endogenous dynamics as well as the strong Markov property of the model seem unrealistic. Although the first can be addressed rather easy with more complicated state transition probabilities which consider also the asset price, the Markov property is critical for the convergence to a diffusion process. More complicated microscopic models leading to non-markovian limits, e.g. solutions of stochastic differential delay equations (see Arriojas et al. \cite{arr2007}), could be investigated in the future, although far less literature in form of limit theorems is available for this case.

\newpage
\section{Appendix}

\subsection{Proof of Proposition \ref{large_market_approx_Q}}\label{proof_prop_large_market_Q}
We apply Theorem 3.6 of Henkel \cite{hen2016}. Note that our model fits in the framework of Henkel \cite{hen2016} with the market character and market charecter index, which are defined in Definition 2.3 and 2.16 in Henkel \cite{hen2016}, given by $M_k = (1-\overline{M}_k,\overline{M}_k)$, $V_t^n = (1-Q_t^n,Q_t^n)$ and $d_1=1$. The expected aggregated transition $b_n$ and transition volume $c_n$ (Henkel \cite{hen2016} Definition 3.3 and 3.4) necessary to determine the limit of $V_t^n$, respectively $Q_t^n$, when $n \to \infty$ are given by

 \begin{equation}
\begin{split}
b_n^2(x,v) &= \frac{1}{n} \sum_{a=1}^n \mu_a (\Pi^{2+}_{n,a}(v) - \Pi^{2-}_{n,a}(v))\\
&= \frac{1}{n} \sum_{a=1}^n \mu_a [(1-v_2)\Pi^{2,1}_{n,a}(v_2) - v_2\Pi^{1,2}_{n,a}(v_2)]\\
&= \frac{1}{n} \sum_{a=1}^n \mu_a \left[(1-v_2) (\beta_a \frac{p_a}{2 \gamma^2_a n } + \frac{\eta_a  (1-v_2) v_2^2}{2}) - v_2 (\beta_a \frac{1-p_a}{2 \gamma^2_a n} + \frac{\eta_a (1-v_2)^2 v_2}{2}) \right]\\
&= \frac{1}{n} \sum_{a=1}^n \frac{\beta_a}{2} (p_a-v_2)\\
b_n^1(x,1) &= \frac{1}{n} \sum_{a=1}^n \mu_a (\Pi^{1+}_{n,a}(v) - \Pi^{1-}_{n,a}(v))\\
&= - b_n^2(x,v)
\end{split}
\end{equation}

 \begin{equation}
\begin{split}
(c_n^2(x,v))^2 &= \frac{1}{n^2} \sum_{a=1}^n \mu_a (\Pi^{2+}_{n,a}(v) + \Pi^{2-}_{n,a}(v))\\
&= \frac{1}{n^2} \sum_{a=1}^n \mu_a [(1-v_2)\Pi^{2,1}_{n,a}(v_2) + v_2\Pi^{1,2}_{n,a}(v_2)]\\
&= \frac{1}{n^2} \sum_{a=1}^n \mu_a\left[(1-v_2) (\beta_a \frac{p_a}{2 \gamma^2_a n } + \frac{\eta_a  (1-v_2) v_2^2}{2}) + v_2 (\beta_a \frac{1-p_a}{2 \gamma^2_a n} + \frac{\eta_a (1-v_2)^2 v_2}{2}) \right]\\
&= \frac{1}{n} \sum_{a=1}^n \left[\frac{\beta_a(p_a - 2 v_2 p_a + v_2)}{2 n} + \gamma_a^2 \eta_a (1-v_2)^2 v_2^2 \right]\\
(c_n^1(x,v))^2 &= \frac{1}{n^2} \sum_{a=1}^n \mu_a (\Pi^{1+}_{n,a}(v) + \Pi^{1-}_{n,a}(v))\\
&= (c_n^2(x,v))^2 \\
(c_n^{1,2}(x,v))^2 &= (c_n^{2,1}(x,v))^2 \\
&= -(c_n^2(x,v))^2 
\end{split}
\end{equation}

Now, by Assumption \ref{Assumption_Q} 

\begin{equation}
b_n=(b_n^1,b_n^2) \xrightarrow{n \to \infty} b := \begin{pmatrix} -1 \\ 1 \end{pmatrix} \beta (p-v_2)
\end{equation}

and

\begin{equation}
c_n =  \begin{pmatrix} c_n^{1} & c_n^{1,2} \\ c_n^{2,1} & c_n^{2}  \end{pmatrix} \xrightarrow{n \to \infty} c := \begin{pmatrix} 1 & -1 \\ -1 & 1 \end{pmatrix} \gamma \sqrt{\eta} (1-v_2)v_2.
\end{equation}

So, by Theorem 3.6 of Henkel \cite{hen2016} $V_t^n \xrightarrow{n \to \infty} V_t$, where $V_t$ is the solution of 
\begin{equation}
dV_t = b(V_t)dt + c(V_t)dB_t, V_0=(1-\theta, \theta)
\end{equation}

and hence $Q_t^n \xrightarrow{n \to \infty} Q_t$, where $Q_t$ is the solution of 

\begin{equation}
dQ_t = \beta (p-Q_t)dt + \gamma \sqrt{\eta} (1-Q_t) Q_t dB_t, \ Q_0 = \theta,
\end{equation}

since $Q_t = V^2_t$.

\subsection{Proof of Proposition \ref{Diff_approx}}\label{proof_diff_approx1}
Also here we apply Theorem 3.6 of Henkel \cite{hen2016}. Since the dynamics of $Q_t^n$ do not depend on $X_t^n$, the convergence of $Q_t^n$ and its limit is given by Proposition \ref{large_market_approx_Q}. To also show the convergence of $X_t^n$ and to determine the limit when $n \to \infty$, we calculate the expected aggregated excess demand $z_n$ and the trading volume $\sigma_n$ (see Henkel \cite{hen2016} Definition 3.1 and 3.3).
\begin{equation}
\begin{split}
z_n(x,v) &= n^{-1/2} \sum_{a=1}^n \lambda_a \E [e_a^n(x,v,s)]\\
&= \frac{1}{\sqrt{n}} \left( \sum^n_{a \in \mathbb{F}_n} \bar{\lambda}_a \E \left[\frac{1}{\sqrt{n}}(F-x)\right] + \sum^n_{a \notin \mathbb{F}_n} 0 \right)
\end{split}
\end{equation}

\begin{equation}
\begin{split}
\sigma_n(x,v)^2 &= \frac{1}{n} \sum_{a=1}^n \lambda_a \E [e_a^n(x,v,s)^2]\\
&= \frac{1}{n} \left( \sum^n_{a \in \mathbb{F}_n} \bar{\lambda}_a \E \left[\frac{1}{n}(F-x)^2 \right]+ \sum^n_{a \notin \mathbb{F}_n} \lambda_a \sigma_{\xi}^2 \gamma^2 \eta q^2 (1-q)^2 \right)\\
&= \frac{1}{n} \left( \sum^n_{a \in \mathbb{F}_n} \bar{\lambda}_a \E \left[\frac{1}{n}(F-x)^2 \right]+ \sigma_{\xi}^2 \gamma^2 \eta q^2 (1-q)^2 \left[ (\sum^n_{a \notin \mathbb{F}_n} \bar{\lambda}_a) + \delta nq \right] \right)
\end{split}
\end{equation}

By Assumption \ref{Assumption_Q} and \ref{Assumption_X} we have

\begin{equation}
z_n(x,v) \xrightarrow{n \to \infty}  \bar{\lambda}_F (F-x)
\end{equation}
and
\begin{equation}
\sigma_n(x,v)^2 \xrightarrow{n \to \infty} (\bar{\lambda}_N + C_e q) \sigma_{\xi}^2 \gamma^2 \eta q^2 (1-q)^2
\end{equation}

After realizing that Assumption 3.5 of Henkel \cite{hen2016} is fulfilled, we apply Theorem 3.6 of Henkel \cite{hen2016} and get Proposition \ref{Diff_approx} as a result.

\newpage
\thispagestyle{plain}
\nocite{*}
\bibliography{Literature2}
\bibliographystyle{abbrv}

\end{document}